\documentclass[aps,pre,preprint,superscriptaddress,showpacs]{revtex4-1}

\usepackage{graphicx}
\usepackage{subfigure}
\usepackage{amsmath}

\begin{document}


\title{Bypass Rewiring and Robustness of Complex Networks}


\author{Junsang Park}
\email[]{junsp85@kaist.ac.kr}
\affiliation{Graduate School of Information Security, Korea Advanced Institute of Science and Technology, 291 Daehak-ro, Yuseong-gu, Daejeon 34141, Republic of Korea}
\author{Sang Geun Hahn}
\affiliation{Graduate School of Information Security, Korea Advanced Institute of Science and Technology, 291 Daehak-ro, Yuseong-gu, Daejeon 34141, Republic of Korea}
\affiliation{Department of Mathematical Sciences, Korea Advanced Institute of Science and Technology, 291 Daehak-ro, Yuseong-gu, Daejeon 34141, Republic of Korea}


\date{\today}

\begin{abstract}
A concept of bypass rewiring is introduced and random bypass rewiring is analytically and numerically investigated with simulations.
Our results show that bypass rewiring makes networks robust against removal of nodes including random failures and attacks.
In particular, random bypass rewiring connects all nodes except the removed nodes on an even degree infinite network and makes the percolation threshold $0$ for arbitrary occupation probabilities.
In our example, the even degree network is more robust than the original network with random bypass rewiring while the original network is more robust than the even degree networks without random bypass.
We propose a greedy bypass rewiring algorithm which guarantees the maximum size of the largest component at each step, assuming which node will be removed next is unknown.
The simulation result shows that the greedy bypass rewiring algorithm improves the robustness of the autonomous system of the Internet under attacks more than random bypass rewiring.
\end{abstract}

\pacs{89.75.Hc, 64.60.ah, 05.10.-a}

\maketitle

\section{INTRODUCTION}

Many real world systems (the Internet, electric power grids, the World Wide Web, social networks, urban streets, airline routes, subway, and others) are represented by complex networks with many nodes and many links between nodes \cite{albert00, cohen00, cohen01, newman01, dorogovtsev02, latora02, holme02, albert02, newman03, boccaletti06, crucitti06, schneider11, newman}.
A network (graph) breaks into small disconnected parts when nodes are deleted.
Complex networks are robust against random failures or errors (random removal of nodes) but fragile and vulnerable to (intentional) attacks (targeted removal of nodes in decreasing order of degree from the highest degree) \cite{albert00, callaway00, cohen00, cohen01, holme02, dorogovtsev02, albert02, newman03, gallos05, boccaletti06, dorogovtsev08, newman}.
There are various mitigation methods which make networks more robust \cite{beygelzimer05, xiao10, schneider11, quattrociocchi14}.
However, there are geographical, economic, and technical problems to implement the mitigation methods known so far.
Therefore, we propose a concept of bypass rewiring to make networks robust against random failures and attacks.

\section{A CONCEPT OF BYPASS REWIRING}

\begin{figure}
\subfigure[]{
	\includegraphics[]{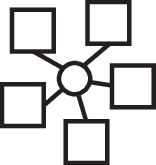}
	\label{fig:before_removal}
}
\subfigure[]{
	\includegraphics[]{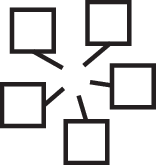}
	\label{fig:without_bypass}
}
\subfigure[]{
	\includegraphics[]{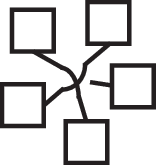}
	\label{fig:with_bypass}
}
\caption{\label{fig:bypass_rewiring} A circle is for a node and squares are for components. (a)~Before removal of the node, one node and five components are connected. (b)~After removal of the node, the network fragments into five smaller components without bypass rewiring. (c)~After removal of the node, the network fragments into two larger components and one smaller component with bypass rewiring.}
\end{figure}

A node in Fig.~\ref{fig:before_removal} is removed by random failures or attacks and turns into the removed node in Fig.~\ref{fig:without_bypass}. Bypass rewiring is to directly connect each pair of links of the removed node like Fig.~\ref{fig:with_bypass}.
Each pair of links for rewiring can be chosen in various ways including random selection like random bypass rewiring and heuristic methods like greedy bypass rewiring algorithm.
If the degree of the removed node is odd, one link remains open.
For example, an engineer or equipment can simply rewire cables (links) of a router (node) on the Internet (network) and relay (and sometimes amplify) the signals directly when the router does not work under random failures or attacks.
Since repair is generally harder than rewiring, bypass rewiring would be a more useful and simpler way to improve the connectivity of the other part of the network except the broken router while the router does not work or is under repair.

\section{RANDOM BYPASS REWIRING}

In this paper, we use generating functions based on the generating function formalism introduced in Refs.~\cite{newman01, callaway00, newman, wilf}.
We define
\begin{eqnarray}
G_0(x) & = & \sum_{k=0}^{\infty}{p_k x^k} \text{,} \label{eq:G_0} \\
G_1(x) & = & \frac{\sum_{k=1}^{\infty}{k p_k x^{k-1}}}{\sum_{k=1}^{\infty}{kp_k}} = \sum_{k=0}^{\infty}{q_k x^k} \text{,} \label{eq:G_1} \\
H_1(x) & = & \sum_{k=0}^{\infty}{h_k x^k} \text{,} \label{eq:H_1}
\end{eqnarray}
where $p_k$ is the probability that a randomly chosen node has degree $k$ and $h_k$ is the probability that a randomly chosen link reaches a small component which has $k$ nodes.
In Eq.~(\ref{eq:G_1}), $q_k$ is the probability that a randomly chosen link reaches a node with degree $k+1$. Since nodes of the giant component do not belong to any small component which has a fixed number of nodes on an infinite network, the probability that a randomly chosen node belongs to the giant component is
\begin{eqnarray}
S & = & \sum_{k=0}^{\infty} {p_k \phi_k \{1 - [H_1(1)]^k\} } = \sum_{k=0}^{\infty} {p_k \phi_k (1 - u^k) } \text{,} \label{eq:S}
\end{eqnarray}
for
\begin{eqnarray}
H_1(x) & = & \sum_{k=0}^{\infty} { q_k \{ 1 - \phi_{k+1} + \phi_{k+1} [H_1(x)]^k \} } \text{,} \label{eq:H_1_H_1} \\
H_1(1) & = & u = f_1(u) = \sum_{k=0}^{\infty} { q_k (1 - \phi_{k+1} + \phi_{k+1} u^k) } \text{,} \label{eq:H_1_u}
\end{eqnarray}
where $\phi_k$ is the occupation probability that a randomly chosen node with degree $k$ is not removed and $u$ is the smallest non-negative real solution of Eq.~(\ref{eq:H_1_u}), that is, $u$ is the average probability that a randomly chosen link is not connected to the giant component \cite{callaway00, cohen01, newman}.
The average occupation probability is
\begin{eqnarray}
\phi = \sum_{k=0}^{\infty}{p_k \phi_k} \text{.} \label{eq:phi_p_k}
\end{eqnarray}

\begin{figure}
\includegraphics[]{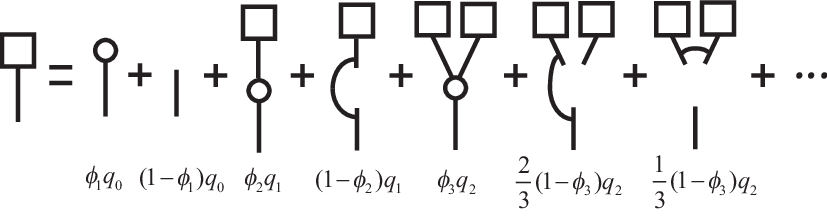}
\caption{\label{fig:gen_bypass} A schematic diagram to calculate the probability that a component (square) is reached by a randomly chosen link with random bypass rewiring under removal of a node (circle).}
\end{figure}

Based on the idea seen in Fig.~\ref{fig:gen_bypass}, $H_1(x)$ and $u$ satisfy
\begin{eqnarray}
H_1(x) & = & q_0 \phi_1 x + q_0 (1 - \phi_1 ) + q_1 \phi_2 x H_1 (x) + q_1 (1 - \phi_2 ) H_1 (x) + q_2 \phi_3 x [H_1 (x)]^2 \nonumber \\
& & + \frac{2}{3} q_2 (1 - \phi_3 ) H_1 (x) + \frac{1}{3} q_2 (1 - \phi_3 ) + \cdots \nonumber \\
& = & \sum_{k=0}^{\infty}{q_k \phi_{k+1} x [H_1 (x)]^k} + H_1(x) \sum_{k=0}^{\infty}{q_k (1 - \phi_{k+1})} \nonumber \\
& & + [1 - H_1(x)] \sum_{k=0}^{\infty} { \frac{p_{2k+1} (1-\phi_{2k+1})}{\sum_{k'=1}^{\infty}{k'p_{k'}}} } \text{,} \label{eq:H_1_H_1_bypass} \\
u & = & f_2(u) = \sum_{k=0}^{\infty}{q_k \phi_{k+1} u^k} + u \sum_{k=0}^{\infty}{q_k (1 - \phi_{k+1})} + (1 - u) \sum_{k=0}^{\infty} { \frac{p_{2k+1} (1-\phi_{2k+1})}{\sum_{k'=1}^{\infty}{k'p_{k'}}} } \text{,} \label{eq:H_1_u_bypass}
\end{eqnarray}
when random bypass rewiring is applied to an infinite network. In the case of random failures ($\phi_k = \phi$), Eq.~(\ref{eq:H_1_u_bypass}) corresponds to
\begin{eqnarray}
u & = & f_3(u) = \phi \sum_{k=0}^{\infty}{q_k u^k} + (1 - \phi) u + (1 - u) (1 - \phi) \sum_{k=0}^{\infty} { \frac{p_{2k+1}}{\sum_{k'=1}^{\infty}{k'p_{k'}}} } \text{.} \label{eq:H_1_u_bypass_failure}
\end{eqnarray}

The self-consistent equations like Eqs.~(\ref{eq:H_1_u}) and (\ref{eq:H_1_u_bypass}) can be solved as follows by the fixed-point iteration, which is a numerical method \cite{burden}.
Iterating
\begin{subequations}
\begin{eqnarray}
u_{i+1} & = & f_1(u_i) \text{,} \label{eq:u_i_f} \\
v_{i+1} & = & f_2(v_i) \text{,} \label{eq:v_i_g}
\end{eqnarray}
\end{subequations}
for $u_0 = v_0 = 0$, 
$u_i$ and $v_i$ approaches to $\bar{u}$ and $\bar{v}$, respectively, as $i$ goes to infinity, for
\begin{subequations}
\begin{eqnarray}
\bar{u} = f_1(\bar{u}) \text{,} \label{eq:u_f} \\
\bar{v} = f_2(\bar{v}) \text{.} \label{eq:v_g}
\end{eqnarray}
\end{subequations}
Since the right-hand side of Eq.~(\ref{eq:H_1_u}) is equal to or larger than the right-hand side of Eq.~(\ref{eq:H_1_u_bypass}) for $0 \leq u \leq 1$,
\begin{eqnarray}
u_i \geq v_i \text{,} \label{eq:u_i_v_i}
\end{eqnarray}
is satisfied for all $i$.
Therefore, $S$ with random bypass rewiring is always equal to or larger than without random bypass rewiring; that is, the percolation threshold with random bypass rewiring is always equal to or smaller than without random bypass rewiring.

To simulate attacks, a node with the highest degree is firstly removed and nodes are removed one by one in decreasing order of degree while randomly chosen nodes are removed one by one in case of random failures.
In the simulation, the degree of each node is not recalculated while nodes are removed.
To simulate random bypass rewiring, each pair of links of the removed node are randomly chosen and rewired until no or one link remains.

\begin{figure}
\subfigure[]{
	\includegraphics[]{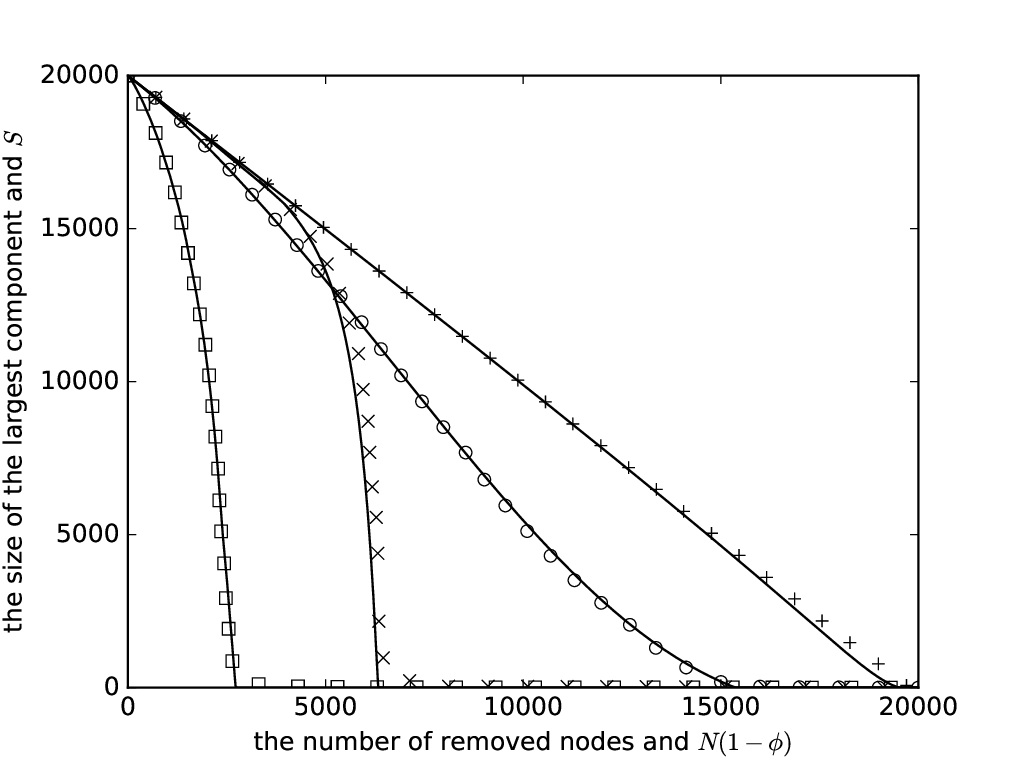}
	\label{fig:sim_num}
}
\end{figure}

\begin{figure}
\subfigure[]{
	\includegraphics[]{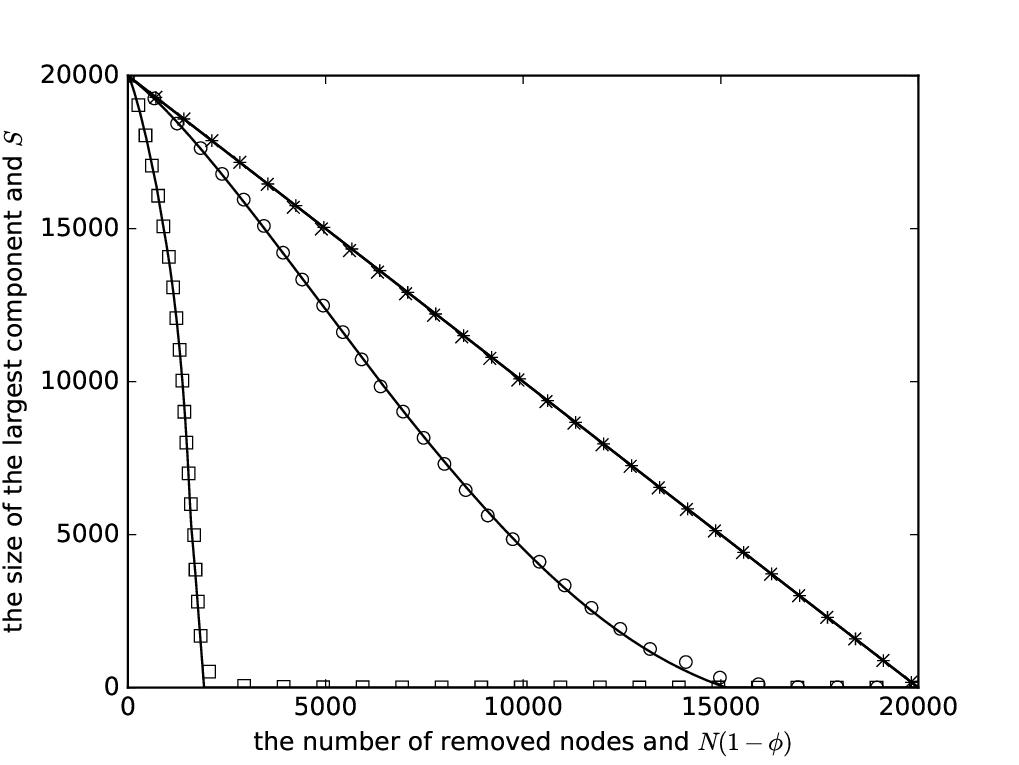}
	\label{fig:sim_num_even}
}
\caption{\label{fig:sim} The size of the largest component with respect to the number of removed nodes under random failures [circles (plus signs)] and attacks [squares (crosses)] without (with) random bypass rewiring. The solid lines are for numerically calculated $S$ with respect to $N(1- \phi)$ from Eqs.~(\ref{eq:S}), (\ref{eq:H_1_u}), (\ref{eq:H_1_u_bypass}), and (\ref{eq:H_1_u_bypass_failure}) on an infinite network with the same degree distribution. (a)~On the undirected scale-free network randomly generated by the configuration model with degree distribution $p_k \sim k^{-3}$, $N = 20000$ nodes, and $M = 30719$ links. (b)~On the undirected even degree scale-free network randomly generated by the configuration model with degree distribution $p'_{2k} = p_{2k} + p_{2k+1}$, $N = 20000$ nodes, and $M=28160$ links. Two straight lines for random bypass rewiring are overlapped.}
\end{figure}

For
\begin{eqnarray}
p_{2k+1} = 0 \text{,} \label{eq:p_2k_1}
\end{eqnarray}
the smallest non-negative real solution of Eq.~(\ref{eq:H_1_u_bypass}) is $u=0$ since the last term of the right side of Eq.~(\ref{eq:H_1_u_bypass}) is $0$.
The smallest non-negative real solution of Eq.~(\ref{eq:H_1_u_bypass_failure}) is also $u=0$ for the same reason.
Therefore, $S$ is equal to $\phi$, and the percolation threshold is $0$ on an even degree infinite network with random bypass rewiring for arbitrary $\phi_k$.
In other words, even degree networks randomly generated are extremely robust against removal of nodes including random failures and attacks with random bypass rewiring.
Figure~\ref{fig:sim_num_even} shows that almost all the nodes except the removed nodes on the even degree network are connected by random bypass rewiring.
Every percolation threshold with random bypass rewiring in Fig.~\ref{fig:sim_num_even} is $0$, while every percolation threshold in Fig.~\ref{fig:sim_num} is not.

The even degree network for Fig.~\ref{fig:sim_num_even} is randomly generated by degree distribution $p'_{2k} = p_{2k} + p_{2k+1}$ where $p_k$ is the degree distribution of the original network for Fig.~\ref{fig:sim_num}.
For this reason, the original network has more links and larger average degree than the even degree network has.
Without random bypass rewiring, the size of the largest component and $S$ on the original network is larger than on the even degree network, respectively.
On the other hand, with random bypass rewiring, the size of the largest component and $S$ on the even degree network are larger than on the original network, respectively, as seen in Fig.~\ref{fig:sim}.
In other words, the even degree network is more robust than the original network with random bypass rewiring, while the original network is more robust than the even degree network without random bypass rewiring.

\section{GREEDY BYPASS REWIRING}

We propose a greedy bypass rewiring algorithm to improve robustness of networks against removal of nodes including random failures and attacks.
The algorithm chooses a pair of link, based on the number of the links not yet rewired and the size of the neighboring components.

A removed node with degree $k$ has $k$ neighbor nodes (neighbor $1$, neighbor $2$, $\dots$, neighbor $k$) and $k$ links (link $1$, link $2$, $\dots$, link $k$).
$R_i$ denotes whether link $i$ is rewired ($R_i=1$) or not yet rewired ($R_i=0$).
Initially, set $R_i=0$ for all $i$ at each step.
$T_{i,j}$ denotes whether neighbor $i$ and neighbor $j$ belong to the same component ($T_{i,j}=1$) or do not ($T_{i,j}=0$), that is, there exists a path from neighbor $i$ to neighbor $j$ without going through the removed node or does not.
$T_{i,i}=1$ is trivially satisfied for all $i$, and
\begin{eqnarray}
Q_{i} = \sum_{j=1}^{k}{T_{i, j} (1-R_j)} \label{eq:Q_i_j}
\end{eqnarray}
denotes how many links in the component to which neighbor $i$ belongs are not yet rewired.
$S_i$ denotes the size of the component to which neighbor $i$ belongs.
At $t$-th step for $1 \leq t \leq \lfloor \frac{k}{2} \rfloor$, choose $\alpha'$ which satisfies $R_{\alpha'} = 0$ and $Q_{\alpha'} \geq Q_i$ with $R_i=0$ for all $i$.
From chosen $\alpha'$, choose $\alpha$ which satisfies $R_\alpha = 0$, $Q_\alpha = Q_{\alpha'}$, and $S_\alpha \geq S_i$ with $R_i=0$ and $Q_i = Q_{\alpha'}$ for all $i$.
Update $R_\alpha=1$.
If $T_{i,j} = 1$ is satisfied for all $i$ and $j$, choose randomly $\beta$ which satisfies $R_{\beta}=0$ without choice of $\beta'$.
Otherwise, choose $\beta'$ which satisfies $R_{\beta'} = 0$, $T_{\alpha, \beta'} = 0$, $Q_{\beta'} \geq Q_i$ with $R_i=0$ and $T_{\alpha,i}=0$ for all $i$.
From chosen $\beta'$, choose $\beta$ which satisfies $R_\beta = 0$, $T_{\alpha, \beta} = 0$, $Q_\beta =Q_{\beta'}$, and $S_\beta \geq S_i$ with $R_i=0$, $T_{\alpha,i}=0$, and $Q_i = Q_{\beta'}$ for all $i$.
Update $R_\beta = 1$.
When neighbor $\alpha$ and neighbor $\beta$ do not belong to the same component ($T_{\alpha, \beta} = 0$), update the size of the component to which neighbor $\alpha$ and neighbor $\beta$ belong, that is, $S_i = [1 - (1-T_{\alpha,i}) (1-T_{\beta,i})] (S_\alpha + S_\beta)$ for all $i$.
Update $T_{i,j} = T_{j,i} = 1$ if there exist $i$ and $j$ which satisfy $T_{i,\alpha} T_{\beta,j} = 1$, that is, $T_{i,j} = T_{j,i} = 1 - (1 - T_{i,j}) (1 - T_{i,\alpha} T_{\beta,j})$.
Repeat each step of the algorithm $\lfloor \frac{k}{2} \rfloor$ times whenever a node is removed.


If there exists $i$ ($\neq \alpha, \beta$) which satisfies $Q_i > 1$, the maximum size of the largest component is not guaranteed for $Q_\alpha \leq 1$ or $Q_\beta \leq 1$.
From this aspect, we choose $\alpha'$ and $\beta' \neq \alpha$ which maximize $Q_{\alpha'}$ and $Q_{\beta'}$ in the algorithm.
If $Q_i \leq 1$ is satisfied for all $i$ ($\neq \alpha$), the maximum size of the largest component is not guaranteed when there exists $i$ which satisfies $S_\beta < S_i$.
If $Q_i \leq 1$ is satisfied for all $i$, the maximum size of the largest component is not guaranteed when there exists $i$ which satisfies $S_\alpha < S_i$ or $S_\beta < S_i$.
From this point of view, we choose $\alpha$ and $\beta$ ($\neq \alpha$) which maximize $S_\alpha$ and $S_\beta$ in the algorithm for $Q_\alpha=Q_{\alpha'}$ and $Q_\beta=Q_{\beta'}$.
Therefore, the algorithm guarantees the maximum size of the largest component at each step where which node will be removed next is unknown.

\begin{figure}
\subfigure[]{
	\includegraphics[]{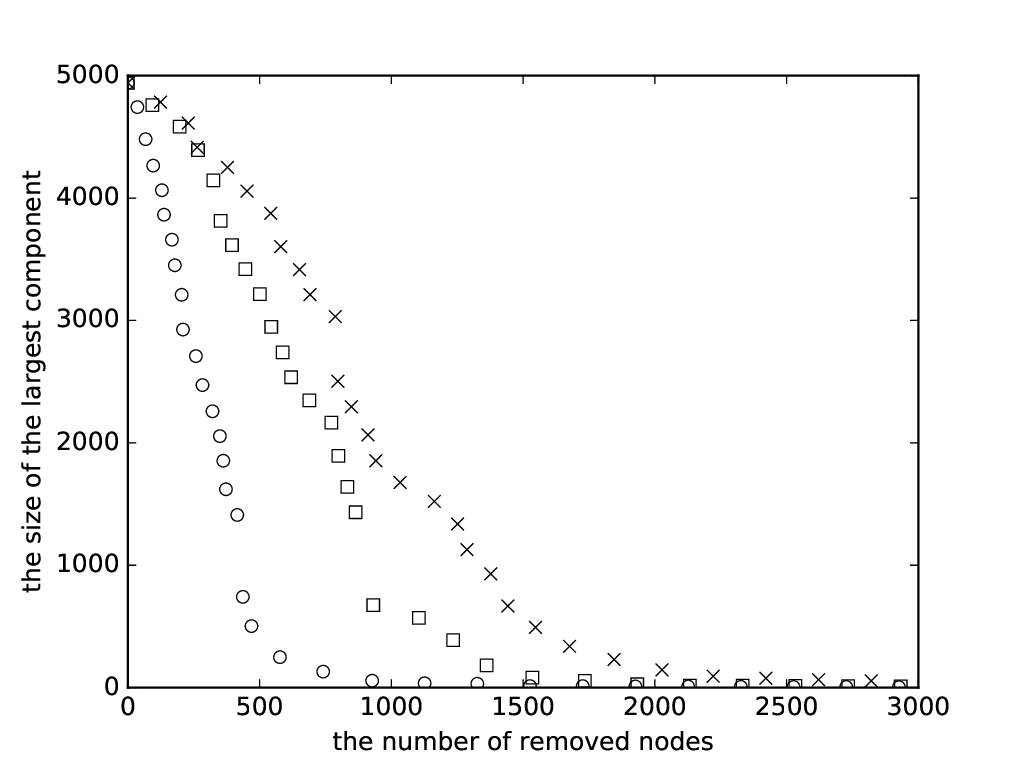}
	\label{fig:power_attack}
}
\caption{\label{fig:internet_power_attack} The circles are for the case without bypass rewiring. The squares (crosses) are for the case with random (greedy) bypass rewiring. (a)~The size of the largest component with respect to the number of removed nodes on the autonomous system (AS-733) from \cite{snapnets} with $N = 6474$ nodes and $M = 12572$ links under attacks. (b)~The size of the largest component with respect to the number of removed nodes on the electrical power grid of the western United States from \cite{watts98} with $N = 4941$ nodes and $M = 6594$ links under attacks.}
\end{figure}

Figure~\ref{fig:internet_power_attack} shows that the greedy bypass rewiring algorithm improves the robustness of the  Internet and the electrical power grid under attacks more than random bypass rewiring.
Since we ignore and eliminate self links and double links for the simulation, the number of links on the Internet is 12572 where the network originally has 13895 links.

\section{CONCLUSIONS}

In summary, we have introduced a concept of bypass rewiring and analytically and numerically investigated random bypass rewiring with simulations.
The results have shown that random bypass rewiring improves robustness of networks under removal of nodes including random failures and attacks.
With random bypass rewiring, all nodes except the removed nodes on an even degree infinite network are connected for arbitrary occupation probabilities, and then the percolation threshold is $0$.
With (without) random bypass rewiring, the size of the largest component and $S$ on the original network are smaller (larger) than on the even degree network randomly generated by the degree distribution $p'_{2k} = p_{2k} + p_{2k+1}$, respectively, where $p_k$ is the degree distribution of the original network.
This means that random bypass rewiring makes even degree networks extremely robust.
Based on the number of the links not yet rewired and the size of the neighboring components, we have proposed a greedy bypass rewiring algorithm which guarantees the maximum size of the largest component at each step, assuming that which node will be removed next is unknown.
The simulation result has shown that the algorithm improves robustness of the autonomous system of the Internet more than random bypass rewiring.
We hope that bypass rewiring equipment is implemented and added on the existing routers on the Internet.
More applications of various kinds and studies of bypass rewiring in many fields are expected.

\bibliography{bypass_rewiring_edited}

\newpage

\begin{figure}
\subfigure[]{
	\includegraphics[]{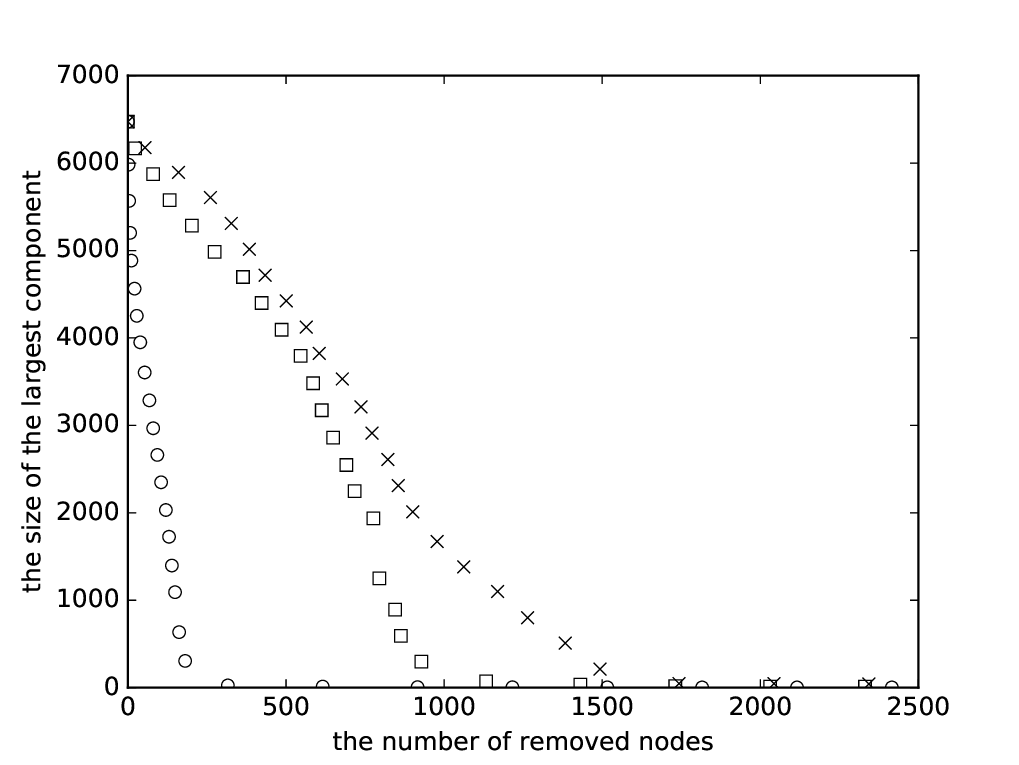}
	\label{fig:internet_attack}
}
\end{figure}

\end{document}